\documentclass[fleqn,usenatbib,useAMS]{mnras}

\usepackage{graphicx}	
\usepackage{amsmath}	
\usepackage{amssymb}	
\usepackage{multicol}        
\usepackage{bm}		
\usepackage{pdflscape}	
\usepackage[table]{xcolor}
\usepackage[flushleft]{threeparttable}

\newcommand{\teff}{$T_{\rm eff}$}

\newcommand{\logg}{$\log g$}

\newcommand{\feh}{$\rm{[Fe/H]}$}

\newcommand{\vmic}{$v_{t}$}

\newcommand{\qq}{$\mathrm{q}^{2}$}

\newcommand{\sm}{$\rm{M_{\odot}}$}
\newcommand{\sr}{$\rm{R_{\odot}}$}

\newcommand{\be}{\begin{equation}}
\newcommand{\ee}{\end{equation}}
\newcommand{\ben}{\begin{eqnarray}}
\newcommand{\een}{\end{eqnarray}}
\newcommand{\bfg}{\begin{figure}}
\newcommand{\efg}{\end{figure}}

\usepackage[T1]{fontenc}
\usepackage{ae,aecompl}

\usepackage{newtxtext,newtxmath}

\title[]{The effect of stellar activity on the spectroscopic stellar
       parameters of the young solar twin HIP 36515}

\author[Yana Galarza et al.]{Jhon Yana Galarza,$^{1}$\thanks{E-mail: ramstojh@usp.br}
Jorge Mel\'endez,$^{1}$
Diego Lorenzo-Oliveira,$^{1}$
Adriana Valio,$^{2}$
\newauthor
Henrique Reggiani,$^{1}$
Mar\'ilia Carlos,$^{1}$
Geisa Ponte,$^{2, 3}$
Lorenzo Spina,$^{4}$
\newauthor
Rapha\"elle D. Haywood$^{5}$\thanks{NASA Sagan Fellow}
and Davide Gandolfi$^{6}$
\\
$^{1}$Universidade de S\~ao Paulo, Departamento de Astronomia do IAG/USP, Rua do Mat\~ao 1226, \
     Cidade Universit\'aria, 05508-900 S\~ao Paulo, SP, Brazil \\
$^{2}$CRAAM, Mackenzie Presbyterian University, Rua da Consola\c{c}\~ao, 896, S\~ao Paulo, Brazil, \\
$^{3}$Universidade Federal do Rio de Janeiro, Observat\'orio do Valongo, Ladeira do Pedro Antonio 43, CEP: 20080-090 Rio de Janeiro,
RJ, Brazil, \\
$^{4}$Monash Centre for Astrophysics, School of Physics and Astronomy, Monash University, VIC 3800, Australia, \\
$^{5}$Harvard-Smithsonian Center for Astrophysics, 60 Garden Street, Cambridge, MA 02138, USA, \\
$^{6}$Dipartimento di Fisica, Universit\`a degli Studi di Torino, via Pietro Giuria 1, I-10125, Torino, Italy
}

\date{Accepted ---. Received ---; in original form ---}

\pubyear{2019}

\begin{document}
\label{firstpage}
\pagerange{\pageref{firstpage}--\pageref{lastpage}}
\maketitle

\begin{abstract}
Spectroscopic equilibrium allows us to obtain precise stellar parameters in Sun-like stars. It relies on the assumption of the iron excitation and ionization equilibrium. However, several works suggest that magnetic activity may affect chemical abundances of young active stars, calling into question the validity of this widely-used method. We have tested for the first time variations in stellar parameters and chemical abundances for the young solar twin HIP 36515 ($\sim$0.4 Gyr), along its activity cycle. This star has stellar parameters very well established in the literature and we estimated its activity cycle in $\sim$6 years. Using HARPS spectra with high resolving power (115 000) and signal-to-noise ratio ($\sim$270), the stellar parameters of six different epochs in the cycle were estimated. We found that the stellar activity is strongly correlated with the effective temperature, metallicity, and microturbulence velocity. The possibility of changes in the Li I 6707.8 \AA\ line due to flares and star spots was also investigated. Although the core of the line profile shows some variations with the stellar cycle, it is compensated by changes in the effective temperature, resulting in a non variation of the Li abundance.
\end{abstract}

\begin{keywords}
stars: solar-type -- 
stars: abundances -- 
stars: activity -- 
stars: atmospheres -- 
stars: fundamental parameters -- 
techniques: spectroscopic
\end{keywords}




\section{Introduction}
Precise stellar parameters (effective temperature \teff, metallicity \feh, surface gravity \logg\ and microturbulence velocity \vmic) are not only important for the characterization of stars themselves and to determine precise chemical abundances, but they are also key for characterizing exoplanets, as the planetary radius or planetary mass depends on the adopted properties of its host star. A common method to determine stellar parameters is through the spectroscopic equilibrium, which is performed by imposing the excitation and ionization balance, and evaluating the lack of dependence of iron abundance with reduced equivalent width. However, other techniques (e.g., via color-temperature calibration and asteroseismology) could be also employed to calculate \teff\ \citep{Casagrande:2010A&A...512A..54C} and \logg\ \citep{Melendez:2006ApJ...641L.133M,Hekker:2013A&A...556A..59H, Campante:2014ApJ...783..123C}. In spite of relatively good agreement between these methods, the results could be quite different for young stars with ages of $\sim$1 Gyr \citep[see Figs. 13 anf 15 of][]{Ramirez:2014A&A...572A..48R}, and the problem may be even worse for younger and more active stars. Observational evidence of oxygen abundance variations (up to $\sim$1 dex) with \teff\ in the Hyades cluster ($\sim$0.6 Gyr) were reported by \citet{Schuler:2006ApJ...636..432S}, perhaps due to an overexcitation effect caused by active regions. \cite{Spina:2014A&A...567A..55S} showed that microturbulence of pre-main sequence stars are greater than those of main-sequence stars, and it could be related to the active chromospheres of young stars \citep{Steenbock:1981A&A....99..192S}, which also have considerable impact on their abundance determination \citep{Spina:2017A&A...601A..70S}.

On the other hand, theoretical studies have demonstrated that magnetic fields could affect the stellar spectra in two ways: through forces on the stellar plasma that change the thermodynamical structure of the atmosphere (indirect effects) and through magnetic Zeeman broadening of the spectral lines (direct effects). Thus, the stellar parameters of young magnetically-active stars could not be as precise as we thought, due to the lack of inclusion of magnetic fields in classical one-dimensional model atmospheres \citep{Castelli:2003IAUS..210P.A20C,Gustafsson:2008A&A...486..951G}. The problems described above could be exacerbated for Sun-like stars younger than about 200 Myr, due to the much higher magnetic field strength in those stars \citep{Rosen:2016A&A...593A..35R}.

Seminal theoretical studies have investigated the impact of magnetic fields on spectral line formation using 1D and 3D model atmospheres \citep{Borrero:2008ApJ...673..470B, Fabbain:2010ApJ...724.1536F,  Fabbian:2012A&A...548A..35F, Moore:2015ApJ...799..150M, Shchukina:2015A&A...579A.112S, Shchukina:2016A&A...586A.145S}. These authors found that several iron lines are sensitive to the aforementioned effects, resulting in variations of $\sim$0.1 dex in abundance. Besides, the impact of these effects on the abundance determinations could be reduced when the direct and indirect effects act in opposite ways. They also found that other elements such as C, O and Si are affected by magnetic fields, varying their abundance by about $\sim$0.02 dex. Furthermore, \cite{Flores:2016A&A...589A.135F} showed observational evidence of modulations ($\sim$2 m\AA) of the equivalent widths (hereafter $EW$) of Fe II lines with the chromospheric cycle of the solar twin HIP 45184 (with $\sim$5.1 yr activity cycle and $\langle S \rangle$ = 0.173). However, \cite{Flores:2018A&A...620A..34F} were not able to find the similar modulations in HD 38858 ($\sim$10.8 yr activity cycle). Precise spectroscopic measurements in the Sun over more than three decades \citep{Livingston:2007ApJ...657.1137L}, show small but measurable variations in some spectral lines (including some strong Fe I lines), but their dependence on the solar activity cycle is not well-established yet. The above studies question the validity of the spectroscopic equilibrium method for determining stellar parameters in young stars.
 
Lithium is an important element that has been used to study potential trends with stellar parameters, chromospheric activity, age and exoplanets \citep[e.g., ][]{DoNascimento:2009A&A...501..687D, Ghezzi:2010ApJ...724..154G, Carlos:2016A&A...587A.100C}. 
Enhancement of the neutral Li I at 6707.8 \AA\ in the presence of sunspots was reported by \cite{Giampapa:1984ApJ...277..235G}. Other studies \citep[e.g., ][]{Pallavicini:1992A&A...253..185P,Fekel:1996IAUS..176..345F, Barrado:1997A&A...326..780B} also found variation of Li $EWs$ with stellar activity in chromospheric active binaries, pre-main-sequence stars and late-type dwarfs in very young clusters. Such enhancements were also observed even in flares of chromospheric active binary stars \citep{Montes:1998A&A...340L...5M}. However, despite these large numbers of studies, the nature of the yield of Li I in starspots and flares is still inconclusive. The aim of this study is, therefore, to investigate for the first time, the influence of magnetic fields along the activity cycle on the stellar parameters and Li abundance of the young solar twin HIP 36515.

\section{Spectroscopic observations and data reduction}
The spectra of the Sun (reflected light from the Vesta asteroid) and HIP 36515 were taken with the HARPS spectrograph at La Silla Observatory (3.6 m telescope) between 2014 to 2019, under the ESO (European Southern Observatory) programs: 192.C-0224(G), 192.C-0224(B), 192.C-0224(C), 0100.D-0444(A), 0101.C-0275(R) and 0103.D-0445. Each spectrum was automatically reduced by the HARPS data reduction software. The spectra have high resolving power $R = \lambda / \Delta \lambda = 115\ 000$ and cover a spectral range 3782$-$6913 \AA. The red and blue part of the spectra were carefully normalized using the \texttt{continuum} task in IRAF, and then combined in order to obtain the highest possible SNR. 

\section{Data Analysis}
\subsection{Stellar Activity Cycle}
\label{sec:stellar_activity_cycle}
\begin{figure}
 \includegraphics[width=\columnwidth]{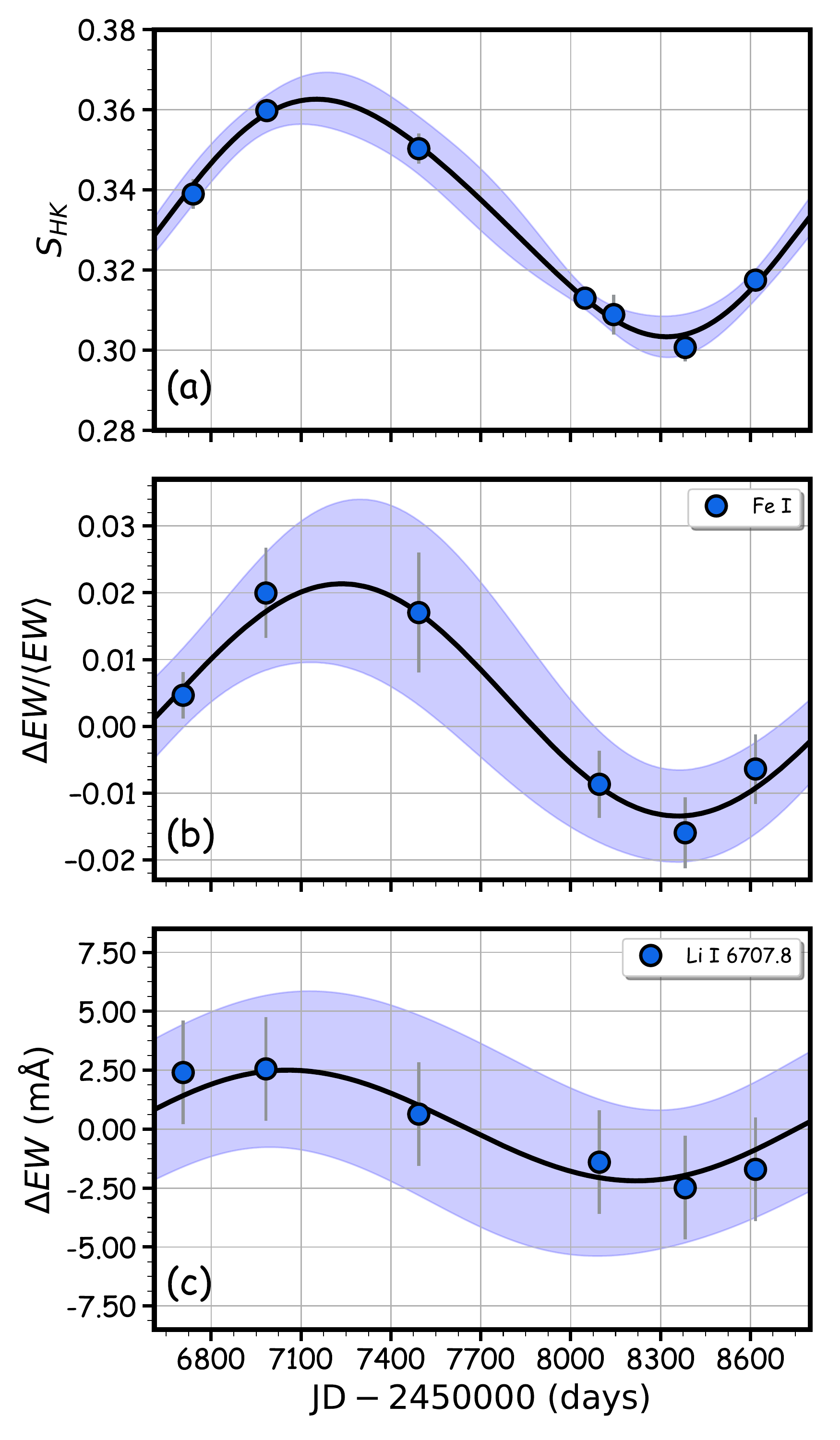}
 \centering
 \caption{\textbf{Panel (a):} Activity cycle fit of HIP 36515 using Gaussian process (black line). Circles represent the $S_{\rm HK}$ values estimated in 7 epochs from 2014 to 2019. \textbf{Panel (b):} Relative changes of $EW$s of our sensitive Fe I lines (Table \ref{tab:1}) along the stellar activity cycle. \textbf{Panel (c):} $\Delta EW$ variation of Li I (6708 \AA) estimated via integrated flux. The black line in panel (b) and (c) represent the predicted changes generated by the activity kernel anchored on $S_{\rm HK}$ index. The shaded regions in all panels are the $2\sigma$ activity variability prediction band. The standard deviation of the residuals computed from the measured and the GP activity induced spectroscopic predictions are 0.0019 and 0.6 m\AA\ for panel (b) and (c), respectively.}
 \label{fig:fig1}
\end{figure}

The stellar and fundamental parameters (\teff, \feh, \logg, age, mass, and radius) of HIP 36515 were estimated with extremely high precision by \cite{Spina:2018MNRAS.474.2580S}. This young active star ($\sim$0.4 Gyr) is an excellent candidate to explore the effects of magnetic fields on abundance determinations. The HARPS-$S$ index (hereafter $S_{\rm HK}$) was estimated measuring the chromospheric emission in the cores of the Ca II H and K lines ($\langle S_{\rm HK} \rangle$ = 0.3270 $\pm$ 0.0226), following the prescriptions presented in \cite{Diego:2018A&A...619A..73L}. In panel (a) of Fig. \ref{fig:fig1} is shown the activity cycle of HIP 35615 binned in 90 days. Inspecting our activity time-series, we estimated the $S_{\rm HK}$ amplitude variations induced by stellar rotation \citep[$\sigma_{S_{\rm HK,Prot}}=0.0046$ and $P_{\rm rot}=4.6\pm1.2$ days;][]{ Diego:2019MNRAS.485L..68L}. We propagated the activity variations within each night of observations ($\sigma_{ S_{\rm HK, night}}=0.0007$) with the errors caused by rotational modulation weighted by the number of observing nights within each bin of 90 days ($N_{\rm nights}$): $\sigma_{S_{\rm HK, bin}}\equiv \left(\left [(\sigma_{S_{\rm HK,Prot}}^2)/N_{\rm nights} + \sigma_{ S_{\rm HK, night}}^2\right]\right)^{1/2}$. The cycle period ($P_{\rm cycle}$) is estimated through Gaussian process (GP) fitting to account the quasi-periodic trends of stellar activity, represented in this case by $S_{\rm HK}$. To do so, we choose an appropriate kernel function that combines white noise term ($\sigma^2\delta_{i,j}$) and exponential sine squared kernel which encapsulates important variables related to periodicity ($f\equiv1/P_{\rm cycle}$), amplitude ($\mathcal{A}$), and the harmonic nature ($\Gamma$) of the time series. The adopted kernel function $k(t_i,t_j)$ relates the covariance between two different epochs $t_i$ and $t_j$: $k(t_i,t_j)\equiv \mathcal{A}\exp\left(-\Gamma\sin^2\left[\pi f|t_i-t_j|\right] \right)+\sigma^2\delta_{i,j}$. All GP hyperparameters (GPH) are determined simultaneously using Markov chain Monte Carlo method (MCMC) optimization \citep{Rasmussen:2006gpml.book.....R,Ambikasaran:2015ITPAM..38..252A} resulting in the following median and 16\% confidence interval values of $0.043^{+0.021}_{-0.032}$, $6.2^{+4.3}_{-0.4}$ years, and $0.032^{+0.017}_{-0.012}$, for $\Gamma$, $P_{\rm cycle}$, and $\mathcal{A}_{S_{\rm HK}}$, respectively. These GPH compose the trained activity kernel which will help us to predict the spectroscopic activity induced effects by fixing the periodic GPH ($\Gamma$ and $P_{\rm cycle}$) and leaving only $\mathcal{A}_{\rm variable}$ to scale according to the spectroscopic variable of interest (e.g. $EW$, \teff, \feh, Li, and so on).  

\subsection{Equivalent Widths ($EW$s)}
\label{sub:EW}
If magnetic fields change the shape of spectral lines, this will be reflected on their abundance determination. As discussed above, iron abundance variations ($\sim$0.014 dex) were only predicted in solar simulation spectra under extreme magnetic fields (80$-$200 G). To detect potential variation of spectral lines in HIP 36515, we selected strategically six bins from the activity cycle (see panel (b) of Fig. \ref{fig:fig1}). Several spectra were combined in these bins in order to obtain a SNR$\sim$270. The $EW$ of iron lines \citep[e.g., ][]{Melendez:2014ApJ...791...14M} were  measured along these epochs using the \texttt{splot} task in IRAF, fitting the line profiles with Gaussians. The mean uncertainty of the $EW$ in each epoch is $\sim$0.7 m\AA. Pseudo-continuum regions were obtained following \cite{Bedell:2014ApJ...795...23B} in a window of 6 \AA. The analysis is based on the line-by-line equivalent width technique \citep[e.g.][]{Melendez:2009ApJ...704L..66M, Yana_Galarza:2016A&A...589A..17Y, Spina:2018MNRAS.474.2580S}. 

\begin{figure}
 \includegraphics[width=\columnwidth]{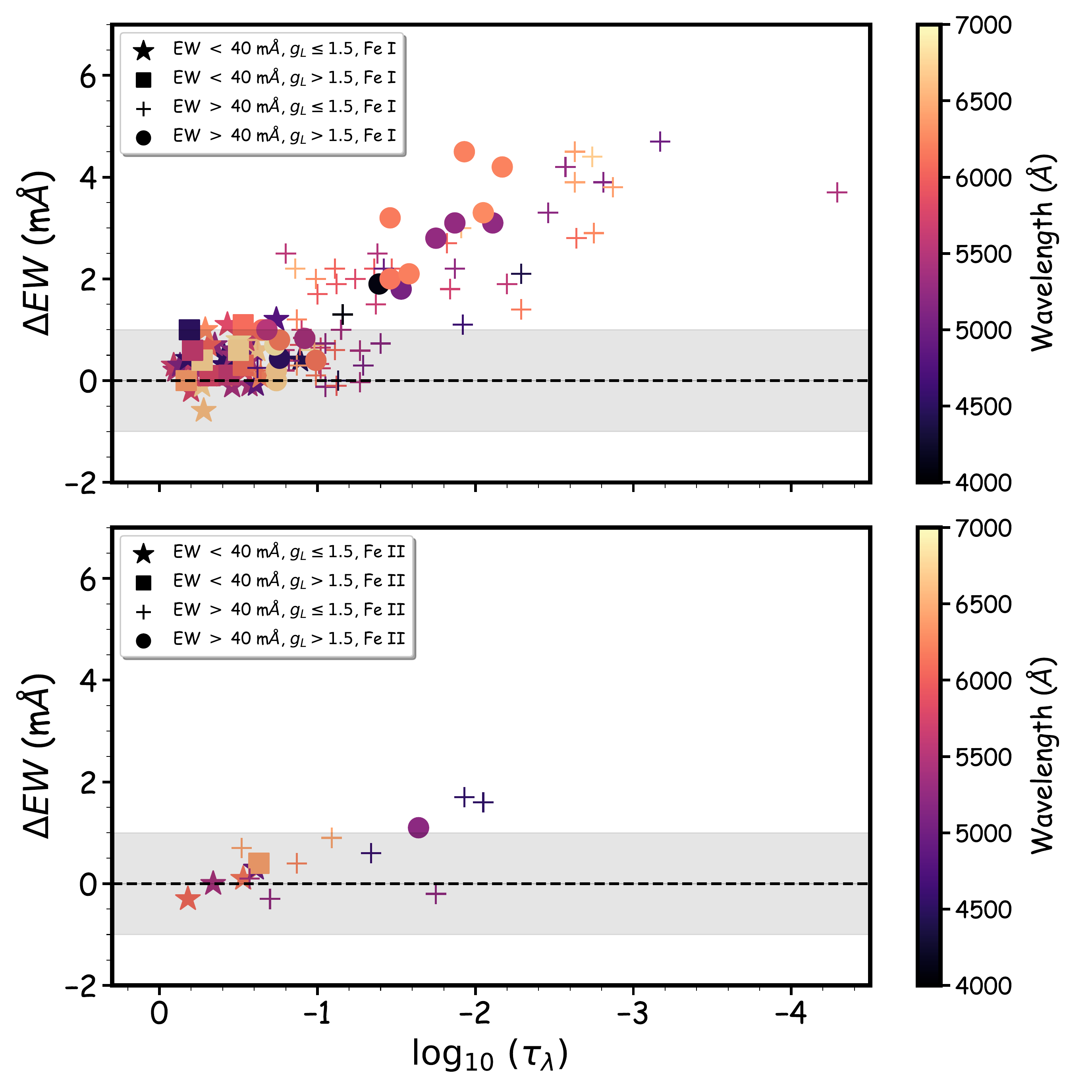}
 \centering
 \caption{Differential $\Delta EW$s of Fe I (upper panel) and Fe II (bottom panel) lines versus their optical depths of formation $\log_{10} (\tau_{\lambda})$. The Fe I and Fe II are divided in four sets, and are colored by their wavelength. Iron lines into the shaded regions are not affected by stellar activity.}
 \label{fig:tau}
\end{figure}

The optical depth ($\tau_{\lambda}$) of the iron lines are estimated by the radiative transfer code MOOG \citep{Sneden:1973PhDT.......180S}, as the depth at the corresponding wavelength and atmospheric layer, using the contribution function given by \citet{Edmonds:1969JQSRT...9.1427E}. Fig. \ref{fig:tau} shows the difference in $EW$s of the iron lines between the maximum and the minimum of the stellar cycle (hereafter $\Delta EW$) versus the mean line-center optical depth formation ($\log_{10} (\tau_{\lambda})$). As in \cite{Shchukina:2015A&A...579A.112S}, we separated the iron lines by $EW$s and land\'e g-factor ($g_{L}$) in four groups. The first two groups have $EW<40$ m\AA\ and are represented by stars ($g_{L} \leq 1.5$) and squares ($g_{L} \geq$ 1.5) in both panels of Fig. \ref{fig:tau}. These lines are formed in regions close to the bottom photosphere ($\log_{10} (\tau_{\lambda})$ > -1), and are into the shaded regions of Fig. \ref{fig:tau}, which represent the 1.5$\sigma$ of the mean uncertainty of the $EW$s, and do not produce relevant impact on the the stellar parameter determinations (only 6 K in \teff, 0.01 dex in \logg, 0.002 dex in \feh, and 0.02 kms$^{-1}$ in \vmic; these values were estimated removing all $\Delta EW > 1.0$ m\AA). The last two sets contain strong and moderate iron lines (40 m\AA\ $<EW<$ 180 m\AA) and are represented by crosses ($g_{L} \leq 1.5$) and circles ($g_{L} \geq 1.5$). There is a clear impact of stellar activity on the $EW$ of Fe I, and it tends to increase with $\log_{10} (\tau_{\lambda})$ (upper panel of Fig. \ref{fig:tau}). The Fe I lines with smaller $\tau_{5}$, meaning those closer to the base of the chromosphere, are the most affected, as they have the largest variations in $\Delta EW$ (Fig. \ref{fig:tau}). We also find a clear correlation between $\Delta EW$ and $g_{L}$ for $EW > 40$ m\AA\ at fixed $\tau_{5}$; however we do not found correlations with another parameters as wavelength, excitation potential and $\log gf$. The bottom panel of Fig. \ref{fig:tau} shows that 3 Fe II lines are sensitive to stellar activity, and we could infer a similar behavior as the Fe I lines. In panel (b) of Fig. \ref{fig:fig1} is shown the relative variations of $EW$s of Fe I lines along the stellar activity cycle. A similar modulation is found for Fe II but with lower amplitude. We also detected $EW$ variations of iron lines found in the literature (see Table \ref{tab:1}).

\subsection{Stellar Parameters}
\label{sub:sp}
We determined the iron abundances (adopting lines only from \cite{Melendez:2014ApJ...791...14M}) using the MOOG code and the Kurucz ODFNEW model atmospheres \citep{Castelli:2003IAUS..210P.A20C}. The stellar parameters along the stellar cycle (i.e., for each bin in panel (b) in Fig. \ref{fig:fig1}) were estimated through the spectroscopic equilibrium using the \textit{qoyllur-quipu} (q$^{2}$) code (see Table \ref{tab:2}). The method was purely differential between the Sun and HIP 36515.

\begin{figure}
 \includegraphics[width=\columnwidth]{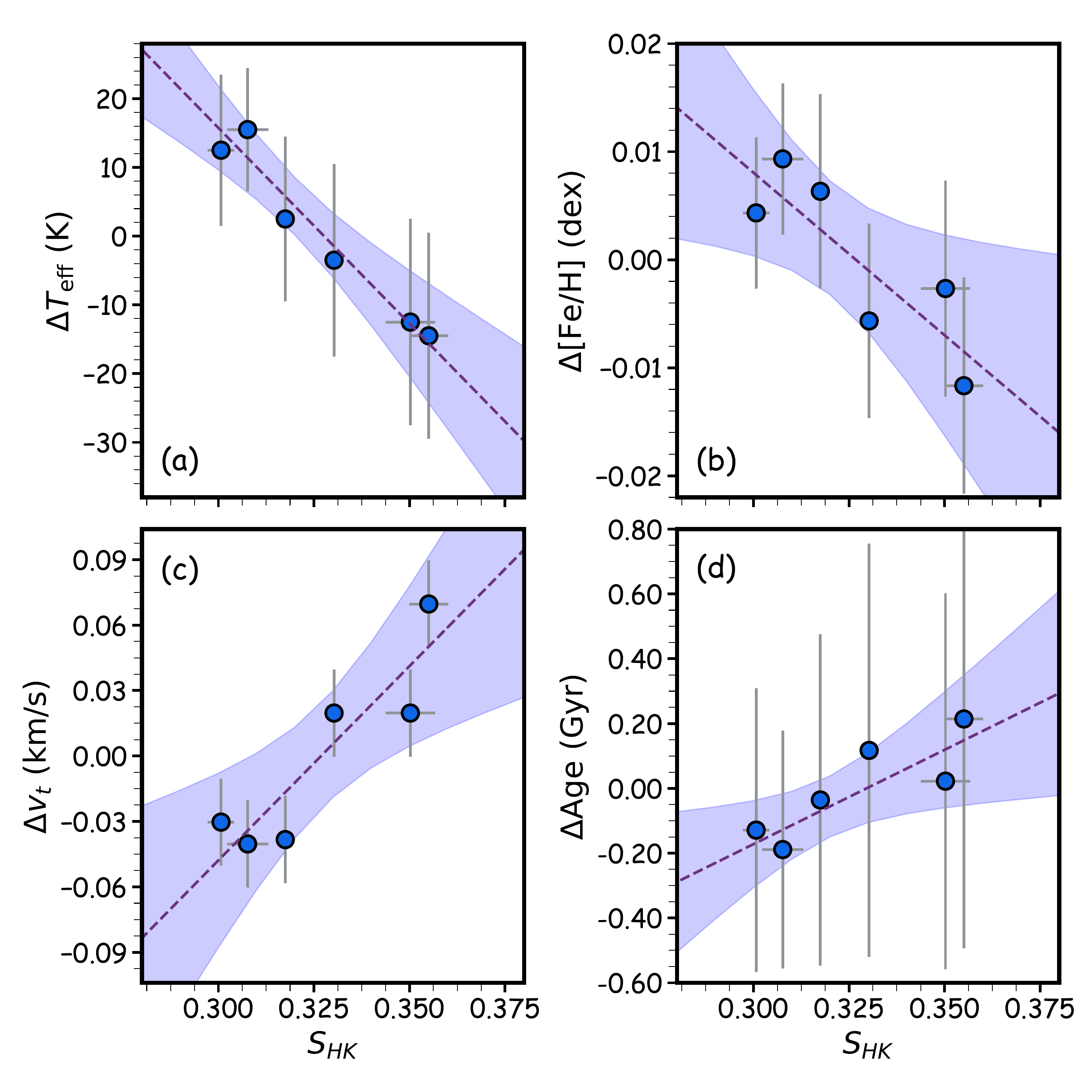}
 \centering
 \caption{Activity induced variations in \teff, \feh, \vmic\ and age. The dashed lines are the best linear fit to the data points, and have been performed using the Kaptein \texttt{kmpfit} package \citep{KapteynPackage}. Shadows regions represent the 95\% confidence interval of the linear fit. The standard deviation of the linear fit for the \teff, \feh, \vmic, and age are 3 K, 0.004 dex, 0.02 kms$^{-1}$ and 0.07 Gyr, respectively.}
 \label{fig:sp_variation}
\end{figure}

Panel (a) of Fig. \ref{fig:sp_variation} shows a fairly strong Spearman anti-correlation ($-$0.94, p-value=0.005) between \teff\ and $S_{\rm{HK}}$. The maximum change observed in \teff\ is 27 K, which is significant considering our high precision of 15 K. From our line list of 104 iron lines, we find that 46 are affected by stellar activity (see Table \ref{tab:1}). These variations impact the iron abundance determinations (maximum variation of 0.016 dex), resulting in an anti-correlation ($-$0.77, p-value=0.07) with $S_{\rm{HK}}$ (see panel (b) of Fig. \ref{fig:sp_variation}). As seen in panel (c) of Fig. \ref{fig:sp_variation}, \vmic\ shows a strong correlation (+0.94, p-value=0.005) with $S_{\rm{HK}}$. The increase of \vmic\ with stellar activity could explain the low \feh\ observed in star forming regions \citep{Spina:2017A&A...601A..70S}; it could be a consequence of changes in the curve of growth due to variations in $EW$s of iron lines.

The ages, masses, and radii were estimated using the \qq\ code through the grid of Yonsei-Yale isochrones \citep{Yi:2001ApJS..136..417Y}.The panel (d) of Fig. \ref{fig:sp_variation} shows no significant changes in the age determination with the activity cycle, since they agree within their uncertainties. Similar results are found for the mass, radius and \logg\ (Table \ref{tab:2}).

\begin{table}
	\centering
	\caption{Iron line list that are sensitive to stellar activity. $^{\rm (a)}$ \citet{Shchukina:2015A&A...579A.112S}, $^{\rm (b)}$ \citet{Fabbian:2012A&A...548A..35F}, $^{\rm (c)}$ \citet{Borrero:2008ApJ...673..470B}. This table is available in its entirety in machine readable format at the CDS.}
	\label{tab:1}
	\begin{tabular}{cccccc} 
		\hline
		\hline
		Wavelength         & $\chi_{\rm exc}$ & $g_{L}$  & $\log_{10} (\tau_{\lambda})$ & $\Delta EW$ & $\langle EW \rangle$ \\
		  (\AA)            &      (eV)        & $\cdots$   &     $\cdots$                   &    (m\AA)   &  (m\AA)              \\
		\hline
		4088.560$^{\rm a}$ &      3.640       &   1.00   &              -1.16                   &     1.3      &     54.5     \\
		4091.560$^{\rm a}$ &      2.830       &   2.00   &              -1.39                   &     1.9      &     59.00    \\
		4389.245           &      0.052       &   1.50   &              -2.29                   &     2.1      &     71.33    \\
		4485.970$^{\rm a}$ &      3.640       &   2.75   &              -0.19                   &     1.0      &     19.50    \\
		4508.288           &      2.855       &   0.50   &              -2.05                   &     1.6      &     89.85    \\
		\hline
		\hline
	\end{tabular}
\end{table}

\begin{table*}
	\centering
	\caption{Stellar and fundamental parameters of HIP 36515 determined in three epochs of the activity cycle. The symbols $^{(*)}$ and $^{(+)}$ represent the stellar parameters obtained using the sensitive and non-sensitive iron lines$^{\alpha}$ (shown in the shadowed region), respectively.}
	\label{tab:2}
	\begin{tabular}{ccccccccc} 
		\hline
		\hline
		Cycle        & $S_{\rm HK}$       & \teff         & \logg            & \feh               & \vmic             & Age             & Mass              & Radius \\
		  $\cdots$     &   $\cdots$           &  (K)          & (dex)            & (dex)              &  (kms$^{-1}$)     & (Gyr)           & \sm               & \sr    \\
		\hline 
		maximum$^{*}$     & 0.355 $\pm$ 0.005 & 5837 $\pm$ 15 & 4.54 $\pm$ 0.027 & -0.045 $\pm$ 0.010 & 1.30 $\pm$ 0.02  & 0.96 $\pm$ 0.70 & 1.033 $\pm$ 0.009 & 0.941 $\pm$ 0.012 \\
		intermediate$^{*}$ & 0.330 $\pm$ 0.002 & 5848 $\pm$ 14 & 4.53 $\pm$ 0.023 & -0.039 $\pm$ 0.009 & 1.25 $\pm$ 0.02  & 0.86 $\pm$ 0.64 & 1.038 $\pm$ 0.009 & 0.943 $\pm$ 0.012 \\
		minimum$^{*}$      & 0.301 $\pm$ 0.004 & 5864 $\pm$ 11 & 4.54 $\pm$ 0.020 & -0.029 $\pm$ 0.007 & 1.19 $\pm$ 0.02  & 0.62 $\pm$ 0.44 & 1.047 $\pm$ 0.005 & 0.945 $\pm$ 0.008 \\
		\rowcolor[HTML]{D6DBDF}
		maximum$^{+}$      & 0.355 $\pm$ 0.005 & 5843 $\pm$ 9 & 4.52 $\pm$ 0.015 & -0.024 $\pm$ 0.007 & 1.14 $\pm$ 0.03  & 0.65 $\pm$ 0.43 & 1.045 $\pm$ 0.006 & 0.945 $\pm$ 0.007 \\
		\rowcolor[HTML]{D6DBDF}
		intermediate$^{+}$ & 0.330 $\pm$ 0.002 & 5838 $\pm$ 8 & 4.51 $\pm$ 0.015 & -0.030 $\pm$ 0.006 & 1.13 $\pm$ 0.03  & 0.90 $\pm$ 0.63 & 1.039 $\pm$ 0.009 & 0.945 $\pm$ 0.010 \\
		\rowcolor[HTML]{D6DBDF}
		minimum$^{+}$      & 0.301 $\pm$ 0.004 & 5843 $\pm$ 9 & 4.52 $\pm$ 0.015 & -0.032 $\pm$ 0.007 & 1.14 $\pm$ 0.03  & 0.70 $\pm$ 0.48 & 1.041 $\pm$ 0.007 & 0.943 $\pm$ 0.008 \\
		\hline
		\hline
	\end{tabular}
	\begin{tablenotes}
	\item $^{\alpha}$ Notice that the list of non-sensitive lines has only two lines at low excitation potential, therefore we suggest caution if this line list is used.
	\end{tablenotes}
\end{table*}

\begin{table}
	\centering
	\caption{Iron line list that is non-sensitive to stellar activity. This table is available in its entirety in machine readable format at the CDS.}
	\label{tab:3}
	\begin{tabular}{cccc} 
		\hline
		\hline
		Wavelength   &     Species     & $\chi_{\rm exc}$ & $\log gf$      \\
		  (\AA)      &     $\cdots$      &      (eV)        &  $\cdots$        \\
		\hline
		4365.896     &      26.0       &      2.990       &  -2.250        \\
		4445.471     &      26.0       &      0.087       &  -5.441        \\
		4523.400     &      26.0       &      3.650       &  -1.960        \\
		4556.925     &      26.0       &      3.250       &  -2.660        \\
		4576.340     &      26.1       &      2.844       &  -2.950        \\
		\hline
		\hline
	\end{tabular}
\end{table}

\subsection{Lithium abundance}
The Li I line at 6707.8 \AA\ is an age indicator in solar twins \citep{Carlos:2016A&A...587A.100C, Carlos:2019MNRAS.tmp..667C}, where younger solar twins are expected to have much larger Li abundances. We analyzed this line along the activity cycle, with the different stellar parameters obtained for each cycle and got very consistent Li abundances. To find the abundances, we fitted synthetic spectra generated with the MOOG code to the observed spectra, and with $\chi^2$ minimization found the LTE abundances of $A(\rm{Li})=2.65$ dex and $A(\rm{Li})=2.66$ dex for the minimum, and maximum of the activity cycle, respectively. A variation of $0.01$ dex is within the expected error of our analysis ($\sigma A(\rm{Li}) = 0.03$ dex) and as such we do not consider this to be indicative of any abundance variation within the activity cycle. 

In order to detect variations in the Li line, we estimated the $EW$ of the synthetic line of the best abundance. We measured the synthetic line, instead of directly measuring the observed spectra, due to blends in the Li line \citep[see Fig. 2 of][]{Carlos:2016A&A...587A.100C}.  Once we found the best Li abundance via synthetic spectra fitting, which accounts for all the blends, we can remove all lines, except the Li lines, and create a clear synthetic lithium profile. We measured the $EW$s of the synthetic lines using  the integrated flux along the Li line profile and found a variation of 4.7 m\AA\ between the maximum and minimum of the activity cycle (see panel (d) of Fig. \ref{fig:fig1}). The standard deviation is $2.25 \ \rm{m}$\AA\ for each measurement. We also estimate the non-LTE abundance via $EW$ employing the grid of abundances from \cite{Lind:2009A&A...503..541L} and found non-LTE abundances variations of only 0.01 dex along the activity cycle, fully compatible with the spectral synthesis analysis. Similar results were obtained when the $EWs$ were measured directly on the observed spectrum using the $splot$ task in IRAF. Thus, our analysis show that although there may be minor variations in $EWs$, they are likely compensated by the changes in {\teff}, so that there are no significant variations on the abundances of Li during the activity cycle, and any minor variations are likely due to measurement errors.

\section{Conclusions}
We have performed, for the first time, a high-precision differential abundance analysis of the young solar twin HIP 35615, along different phases of its activity cycle. We used high-quality HARPS spectra in order to quantify the possible impact of stellar activity on the stellar parameters determination and Li abundance. We determined the activity cycle period ($\sim$6 years) via Gaussian process fitting. Our iron line list has minimal blending and spans a wide range in oscillator strength, Land\'e factor, wavelength, and  the formation height covers a large region of the photosphere (Fig. \ref{fig:tau}). 

We measured the $EW$ of the iron lines in six points of the activity cycle. We report for the first time variations of the $EW$ of several iron lines (43 lines of Fe I and 3 of Fe II) with the stellar activity cycle. These could be used as a new activity indicator for young stars. Despite that these changes are correlated with the Land\'e factor at fixed $\tau_{\lambda}$, the strongest correlation found is with the height of formation of the iron lines (see Fig. \ref{fig:tau}), which could indicate a purely chromospheric effect (i.e. chromospheric heating) over the strong iron lines ($EW>40$ m\AA). This result makes difficult to conclude if these changes are due to the direct or indirect effects.

We also observe that stellar activity influences the stellar parameters determinations based on spectroscopic equilibrium. 
The $EW$ of iron is enhanced during the maximum of the activity cycle, but the impact of activity on the other stellar parameters, results in a decrease of the metallicity. The microturbulence velocity increase with the activity cycle as a consequence of changes in the line profile of iron lines. It is not possible to detect modulations of the surface gravity due to the uncertainties ($\sim$0.027 dex). Other fundamental parameters such as age, mass and radius remain almost invariant within measurement uncertainties.

We created a new line list where the iron lines are slightly sensitive to magnetic fields (Table \ref{tab:3}). Using this list, we determined the stellar parameters in each epoch of the cycle, and we recover similar values obtained in the minimum of the activity cycle, except for the \teff\ (see the shadowed region in Table \ref{tab:2}). It is possible that these values found represent the true stellar parameters of HIP 36515. If spectroscopic observations are available at several phases of the activity cycle, we recommend adopting the stellar parameters determined at the minimum of the activity cycle. 

The $EW$ of the Li I line at 6707.8 \AA\  was measured along the activity cycle, and we found a minor variation of the $EW$s (panel (c) of Fig. \ref{fig:fig1}). However, we found almost no variations of their abundance because the $EW$s are compensated with the \teff\ estimated for each epoch of the activity cycle.

This experiment, made with a solar twin of 0.4 Gyr, should be repeated with even younger solar twins, as the effects could be much stronger, specially for Sun-like stars with ages $<$250 Myr, where the magnetic field strength is significantly higher \citep{Rosen:2016A&A...593A..35R}.

\section*{Acknowledgments}
J.Y.G. acknowledges the support from CNPq. D.L.O. and J.M. thank the support from FAPESP (2016/20667-8; 2018/04055-8). L.S. acknowledge financial support from the Australian Research Council (Discovery Project 170100521). This work was performed under contract with the California Institute of Technology (Caltech)/Jet Propulsion Laboratory (JPL) funded by NASA through the Sagan Fellowship Program executed by the NASA Exoplanet Science Institute (R.D.H.). \\



\bibliographystyle{mnras}
\bibliography{references}







\bsp	
\label{lastpage}
\end{document}